# Zero-Field Optical Readout of Quintet States in Through-Space Coupled Tetracene Oligomers at Room Temperature

Jeannine Grüne[1,2,*], Maciej Majdecki[3], William K. Myers[2], Przemysław Gaweł[3,*], Hannah L. Stern[1]

[1]Department of Materials, University of Oxford, Oxford, UK

[2]Centre for Advanced Electron Spin Resonance (CAESR), University of Oxford, Oxford, UK

[3]Centre of New Technologies, University of Warsaw, Warsaw, Poland

*Corresponding authors: jeannine.grune@materials.ox.ac.uk (J. Grüne), p.gawel@cent.uw.edu.pl (P. Gaweł)

## Abstract

Optically addressable electron spins are promising platforms for quantum technologies, with molecular systems offering chemical control over their spin and optical properties. Achieving optical readout of molecular spin states at room temperature and without an applied magnetic field is an important step towards their practical implementation. Quintet states generated via singlet fission are promising optically addressable multilevel qudits, but their intrinsic spin transitions at zero and weak magnetic fields remain largely unexplored, while substantial room-temperature quintet coherence has predominantly been demonstrated in structurally constrained molecular architectures. Here, we combine zero-field optically detected magnetic resonance (zero-field ODMR) with transient and pulsed electron paramagnetic resonance (EPR) to investigate the magnetic-field dependence and coherent spin properties of quintet states in conformationally flexible, through-space coupled tetracene oligomers. We demonstrate zero-field optical readout of quintet states at room temperature in both tetracene dimers and hexamers, and map the response of optically detected transitions to weak applied magnetic fields. Complementary EPR measurements support quintet formation through stochastic exchange fluctuations and demonstrate a room-temperature phase-memory time of 446 ns, showing that the conformational dynamics enabling high-spin formation remain compatible with substantial spin coherence. Together, these results integrate field-free optical readout, weak-field response and room-temperature spin coherence within a chemically modular, through-space coupled molecular architecture, expanding the design space for optically addressable molecular qudits and quantum sensors.

## Introduction

Molecular spin systems offer chemically tunable platforms for quantum information science, with opportunities in quantum sensing and communication.[1-3] Their molecular nature enables spin properties, local environments and optical pathways for spin readout to be engineered through chemical design.[4,5] However, development of molecular spins for scalable quantum technologies could be facilitated by the ability to initialise, manipulate and read out electron spin states at practical operating conditions, including room temperature and without large applied magnetic field. Operation at zero magnetic field is particularly attractive because it reduces magnetic-field infrastructure while providing direct access to the intrinsic zero-field eigenstates of the molecule.[6,7] Optically detected magnetic resonance (ODMR) probes microwave-driven changes in the spin-state populations through their effect on photoluminescence and can provide optical spin readout even in the absence of an applied magnetic field (zero-field ODMR).[8-10] In solid-state defect centres, such as the nitrogen-vacancy defect in diamond, optical addressability of triplet states at zero field is well established, and the ability to perform ODMR under aligned bias fields of zero to tens of mT is highly advantageous for a range of quantum sensing protocols.[11,12] However, the extension of room temperature zero- and low-field ODMR to molecular systems has emerged only recently, requiring both non-equilibrium spin polarisation and a spin-dependent optical pathway.[13-15] Especially the direct experimental characterisation of higher-spin molecular states at zero-field remains at an early stage.

Quintet states ($S = 2$) are high-spin states with five spin sublevels, providing a multilevel qudit in which information can be encoded across multiple spin states and manipulated through transitions connecting different pairs of states. Singlet fission (SF) provides a route to generate such high-spin states with strong non-equilibrium spin polarisation following photoexcitation. Thereby, a photogenerated singlet exciton is spin-conservatively converted into a strongly correlated triplet pair $^1$TT, which can evolve into a spin-polarised quintet $^5$TT.[6] The reverse process provides a spin-dependent pathway from the triplet-pair manifold back to the emissive singlet, connecting quintet formation to optical readout.[16] SF-derived quintets therefore combine a multilevel spin system with optical initialisation and readout, making them promising candidates for molecular qudits.[17-21]

Room-temperature optical readout of molecular SF spin states has only recently emerged. We previously demonstrated the first room-temperature ODMR of quintet-mediated emission in through-bond diphenylhexatriene oligomers,[22] and analogous optical readout has since been reported in a rigid macrocyclic rubrene dimer.[23] Whilst these studies establish the feasibility of room-temperature optical quintet readout, important questions remain to be addressed. The spin polarisation and optically addressable transitions in the zero-to-low-field regime remain largely unexplored in chemically modular and conformationally flexible SF architectures. First, the spin polarisation of the SF-derived quintets has been characterised predominantly by transient EPR (trEPR) at ~350 mT (X-band) via microwave detection.[24,25] At such fields the Zeeman interaction greatly exceeds the intrinsic zero-field-splitting energy $D_Q$ of the quintet, so that the observed polarisation is resolved in the Zeeman basis. These measurements have established strongly non-equilibrium populations of the quintet $m_s$ sublevels, but do not directly establish the optically accessible transitions of the eigenstates at zero-field. Additionally, substantial room-temperature quintet coherence has so far been demonstrated predominantly in structurally constrained molecular architectures.[23,26] Whether comparable coherence can be retained in a conformationally flexible architecture is therefore an important question for the development of practical SF-based qudits.

Through-space tetracene oligomers based on a hexaphenylbenzene (HPB) scaffold provide a particularly suitable platform to explore zero-field optically addressable molecular quintet states. Here, we study two oligomers, **Tc2** and **Tc6**, in which two or six TIPS-tetracene chromophores are appended to a central HPB core through acetylene linkers, so that inter-chromophore electronic coupling proceeds through space rather than through direct covalent conjugation.[27] The HPB scaffold maintains neighbouring tetracenes in a non-parallel arrangement, while the alkyne-linked chromophores retain local conformational degrees of freedom. Previous transient absorption established efficient intramolecular SF in these molecules, with predominantly strongly bound $^1$TT states and no clear formation of weakly-correlated triplet pairs $^n$(T…T), with n = 1, 3 or 5, or free triplets $T_1$.[27] This makes this system particularly attractive for molecular qudit applications: retention of a strongly correlated triplet pair provides access to the $S = 2$ quintet state, whereas dissociation into independent triplets removes the intrinsic five-level spin system. However, whether these strongly correlated pairs access spin-polarised quintet states, and whether these states can subsequently be optically read out, remained unknown. The chemically modular through-space architecture further provides an opportunity to determine whether the resulting quintet states retain spin coherence at room temperature.

Here, we establish spin-polarised quintet formation, zero-field optical readout and room-temperature spin coherence in through-space coupled tetracene oligomers. Via zero-field and field-dependent ODMR, together with transient and pulsed EPR, we establish the spin properties of the quintets from their intrinsic zero-field spin states through the low-field crossover to their high-field spin polarisation and coherent dynamics. Together, these experiments establish direct optical access to the intrinsic field-free spin states of SF-derived quintets and show that room-temperature coherence can be retained within a conformationally flexible through-space coupled molecular architecture.

## Results

### Zero-field ODMR of room-temperature quintet states

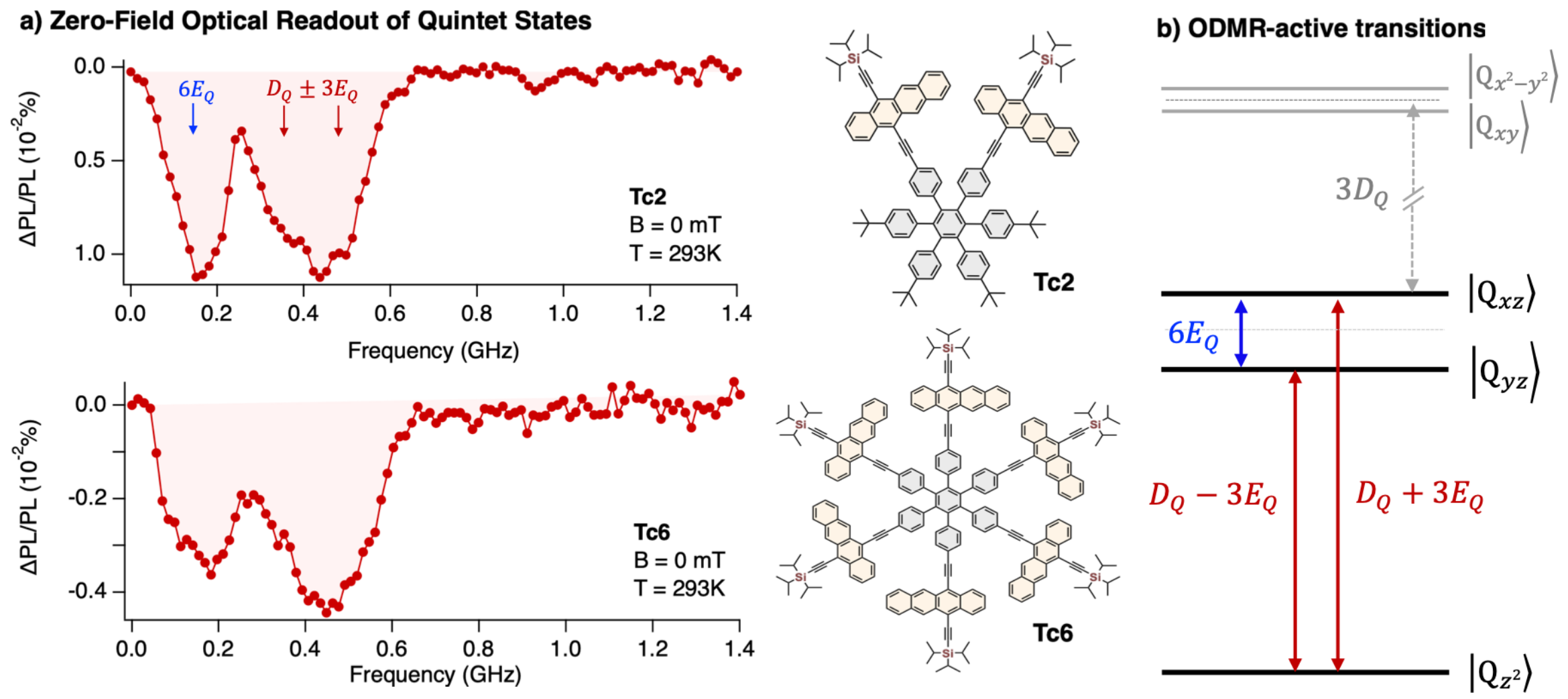

**Figure 1. Room-temperature zero-field ODMR and zero-field spin transitions of Tc2 and Tc6.** a) Room-temperature zero-field ODMR spectra of **Tc2** and **Tc6** in 0.1wt% polystyrene film and corresponding molecular structures. Two negative ΔPL/PL resonances at ~180 MHz ($6E_Q$) and ~450 MHz ($D_Q \pm 3E_Q$) are observed for both oligomers. b) Zero-field quintet energy-level scheme and assignment of the ODMR-active transitions. The observed ODMR-active transitions $D_Q \pm 3E_Q$ and $6E_Q$ are shown in red and blue, respectively; no ODMR transitions are observed towards the upper manifold.

We first probe the quintet state of **Tc2** and **Tc6** directly at zero magnetic field using continuous-wave (cw) ODMR at room temperature. Fig. 1a shows the ODMR contrast ΔPL/PL, defined as the microwave-induced change in photoluminescence, ΔPL, normalised to the total PL. The spectra were recorded under continuous 532 nm optical excitation while sweeping the microwave frequency from 0 to 1.4 GHz, covering the expected zero-field splitting (ZFS) of tetracene-derived triplet-pair states.[18,28] Both oligomers exhibit clear ODMR resonances near 180 and 450 MHz, at frequencies consistent with the expected ZFS of strongly coupled tetracene $^5$TT states. We do not observe ODMR signals attributable to weakly coupled triplet pairs or free triplet states, further supported by the EPR measurements discussed below. Both oligomers exhibit clear ODMR resonances with negative ΔPL/PL contrast, demonstrating that the quintet states can be optically generated, addressed by microwave irradiation and optically read out at room temperature without an applied magnetic field.

The non-equilibrium spin polarisation established following photoexcitation defines the initial populations available for spin manipulation and readout. At zero field, the quintet provides transitions between Cartesian eigenstates that can be selectively addressed at distinct microwave frequencies, allowing access to different two-level subspaces within the multilevel spin system. To assign the observed resonances, we consider the five Cartesian quintet eigenstates at zero field (Fig. 1b): $|Q_{z^2}\rangle$ is the lowest-energy state, followed by $|Q_{yz}\rangle$ and $|Q_{xz}\rangle$ with energy $D_Q$ above, while the upper $|Q_{xy}\rangle$ and $|Q_{x^2-y^2}\rangle$ states lie additionally $3D_Q$ higher. The $|Q_{yz}\rangle$ and $|Q_{xz}\rangle$ states are themselves split by $6E_Q$.[6,16] Considering first **Tc2**, the broad resonance centred near 450 MHz is assigned to the overlapping $|Q_{z^2}\rangle \rightarrow |Q_{yz}\rangle$ and $|Q_{z^2}\rangle \rightarrow |Q_{xz}\rangle$ at $D_Q - 3E_Q$ and $D_Q + 3E_Q$, respectively.[16] Because $E_Q$ is small compared to $D_Q$, these transitions are not resolved individually. The lower-frequency ODMR feature near 180 MHz corresponds to the direct $|Q_{yz}\rangle \leftrightarrow |Q_{xz}\rangle$ transition at $6E_Q \approx 180$ MHz. EasySpin simulations[29] reproduce the observed resonance positions using an $S = 2$ (quintet) spin Hamiltonian with rhombic ZFS parameter $|E_Q| = 30$ MHz and axial ZFS parameter $|D_Q| = 430$-$445$ MHz, the latter consistent with a distribution of local quintet environments within the conformationally flexible through-space coupled system (Fig. S1 and Table S1). The value of $D_Q$ is in agreement with the strong-exchange relation $D_Q \approx D_T/3$, with $D_T$ representing the ZFS parameter for the free triplet, expected for a strongly correlated tetracene triplet pair.[18]

The observed ODMR transitions also provide information on the non-equilibrium population of the zero-field quintet states. An ODMR contrast requires a microwave-drivable population difference between the spin sublevels as well as a difference in the optical pathway of the respective sublevels. Accordingly, the $D_Q \pm 3E_Q$ resonances demonstrate a population difference between $|Q_{z^2}\rangle$ and the $|Q_{yz}\rangle/|Q_{xz}\rangle$ states and are consistent with a preferential population of $|Q_{z^2}\rangle$. In the parallel JDE model of stochastic $^1TT \rightarrow {}^5TT$ formation, population transfer to $|Q_{z^2}\rangle$ scales with $D_T^2$, where $D_T$ is the axial ZFS parameter of the individual triplet, making this a strong quintet-formation pathway.[6,30] The additional $6E_Q$ resonance demonstrates a population difference between $|Q_{yz}\rangle$ and $|Q_{xz}\rangle$. These states are not populated directly from $^1TT$ in the ideal parallel-chromophore limit, whereas extension of the JDE framework to non-parallel chromophores[31] shows that anisotropic inter-chromophore interactions open additional coupling channels where $^1TT$ populates $|Q_{yz}\rangle$ and $|Q_{xz}\rangle$. The observation of the $6E_Q$ transition is therefore consistent with the non-parallel relative orientation of neighbouring tetracene units imposed by the HPB scaffold and provides spectroscopic evidence for quintet-formation pathways beyond the ideal parallel limit.

The zero-field resonances display negative $\Delta PL/PL$ contrast. Within the reverse SF (triplet-triplet annihilation) framework,[16,22] the negative ODMR contrast is consistent with microwave excitation redistributing population away from quintet sublevels that are more strongly connected to the emissive singlet pathway. In particular, $|Q_{z^2}\rangle$ is connected to the singlet manifold through the $^1TT \rightarrow {}^5TT$ formation pathway, retaining a corresponding reverse pathway towards the singlet manifold. Microwave-driven transfer towards states with a weaker connection to the emissive singlet pathway therefore decreases the delayed fluorescence, producing negative ODMR contrast. This intrinsic connection between SF-derived spin polarisation and the reverse pathway to the emissive singlet thus provides a natural mechanism for optical initialisation and readout of the quintet state.

Not all microwave-allowed quintet transitions, however, produce detectable optical contrast. No ODMR feature is observed at $3D_Q \approx 1.3$ GHz for the transitions from the $|Q_{yz}\rangle/|Q_{xz}\rangle$ to the highest $|Q_{xy}\rangle/|Q_{x^2-y^2}\rangle$ pair. Their absence indicates that microwave redistribution involving the upper states does not generate sufficient differential photoluminescence contrast to produce a detectable ODMR response under the present conditions (Supplementary Note 3). The detectable zero-field ODMR response is consequently dominated by transitions within the lower three-state manifold of $|Q_{z^2}\rangle \leftrightarrow |Q_{yz}\rangle/|Q_{xz}\rangle$ and $|Q_{yz}\rangle \leftrightarrow |Q_{xz}\rangle$ as the principal optically addressable part of the quintet at zero field.

**Tc6** exhibits the same zero-field ODMR transitions as **Tc2**, with resonances at the $6E_Q$ and $D_Q \pm 3E_Q$ energies and a comparable overall spectral response (Fig. 1a). Correspondingly, ODMR (Table S1) and trEPR simulations (Table S2) yield similar ZFS parameters for the two oligomers. This indicates that the local spin environment of the quintet-bearing tetracene pair is largely preserved despite extending the molecular architecture from two to six chromophores. The quintet therefore retains the character of a locally coupled tetracene dimer, consistent with SF occurring predominantly between nearest-neighbour tetracene units within the HPB scaffold. The preservation of these local quintet properties across the oligomer series allows the subsequent detailed field-dependent spin characterisation to focus on **Tc2** as the minimal molecular unit.

## Low-field response of optically addressable quintet states

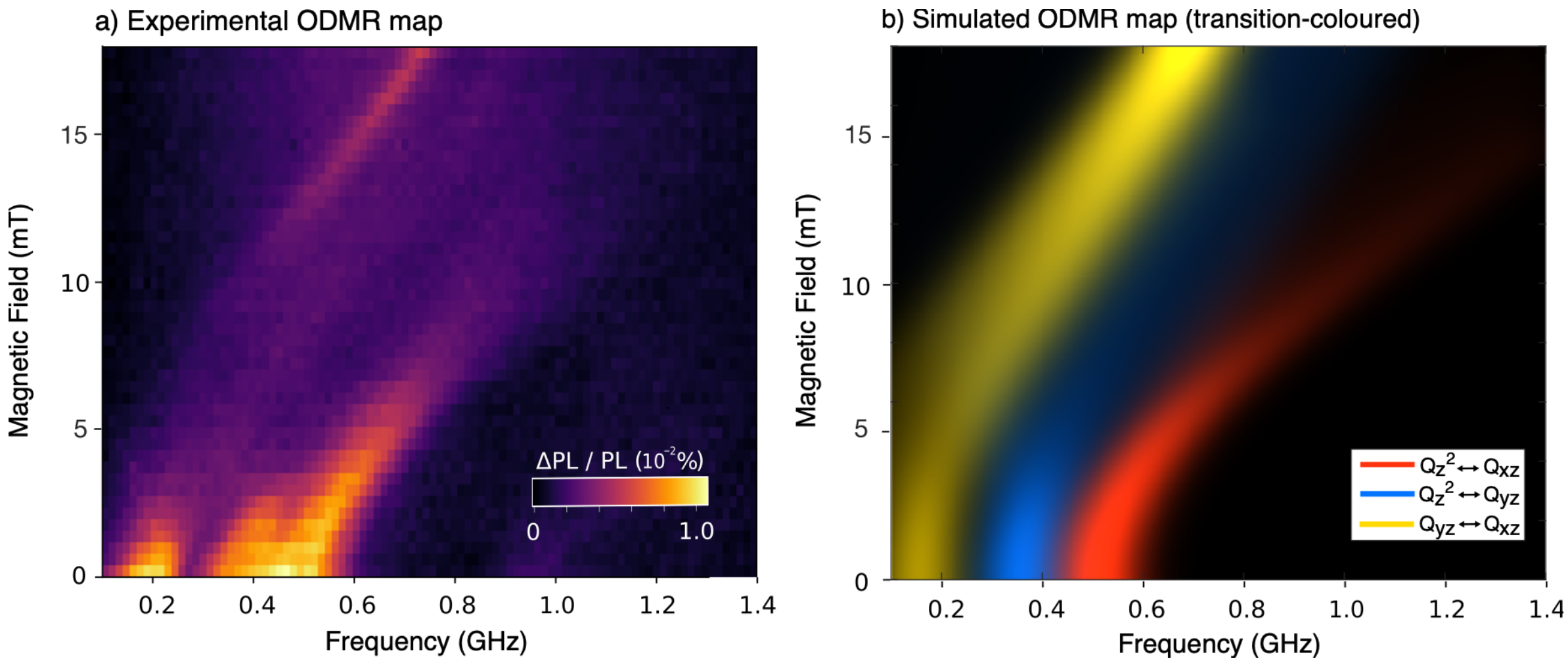

**Figure 2. Field-dependent ODMR of Tc2 from 0 to 18 mT.** a) Experimental 2D ODMR map (total contrast) of **Tc2** at room temperature. b) Corresponding simulated 2D ODMR map with dominant transitions averaged over all three canonical orientations (powder averaged), colour-coded: $|Q_{z^2}\rangle \rightarrow |Q_{xz}\rangle$ (red), $|Q_{z^2}\rangle \rightarrow |Q_{yz}\rangle$ (blue), and $|Q_{yz}\rangle \leftrightarrow |Q_{xz}\rangle$ (yellow).

Having established the optically addressable zero-field transitions, we next examine their response to weak applied magnetic fields, a key characteristic for their potential use in molecular spin sensing. The regime in which the Zeeman interaction becomes comparable to the intrinsic quintet zero-field splitting provides a natural range over which to follow this evolution. Fig. 2a shows a two-dimensional ODMR map of **Tc2** recorded as a function of magnetic field from 0 to 18 mT, tracking the evolution of the zero-field resonances as the Zeeman interaction becomes comparable to the intrinsic quintet ZFS at $D_Q/g_{\mu}B \approx 15$ mT (an analogous 2D map for **Tc6** is shown in Fig. S2). At zero field, the map reproduces the $6E_Q$ transition near 180 MHz and the overlapping $D_Q \pm 3E_Q$ transitions near 450 MHz. With increasing field, the ODMR contrast evolves in both frequency and intensity: the $6E_Q$ feature loses contrast within the first few millitesla, whereas part of the $D_Q$-scale transition persists to higher fields.

Fig. 2b shows the calculated field-dependent transition frequencies for the three dominant transitions identified at zero field: $|Q_{z^2}\rangle \rightarrow |Q_{xz}\rangle$ (red), $|Q_{z^2}\rangle \rightarrow |Q_{yz}\rangle$ (blue), and $|Q_{yz}\rangle \leftrightarrow |Q_{xz}\rangle$ (yellow). The calculated transitions are powder-averaged over the three canonical molecular orientations B‖x, y, z. The calculations show that the different experimental responses arise from their distinct orientational dispersion under an applied magnetic field. The $|Q_{yz}\rangle \leftrightarrow |Q_{xz}\rangle$ and the lower-frequency $|Q_{z^2}\rangle \rightarrow |Q_{yz}\rangle$ branches rapidly disperse in frequency for the different molecular orientations, broadening the ensemble response and causing the corresponding ODMR features to lose contrast within the first few millitesla. This results from a weak Zeeman interaction dispersing their transition frequencies in opposing directions across the three canonical orientations, causing the ODMR transition to broaden and lose contrast in the powder average, rendering them susceptible to even weak millitesla fields. In contrast, the higher-frequency component $|Q_{z^2}\rangle \rightarrow |Q_{xz}\rangle$ at $D_Q + 3E_Q$ retains contrast to higher fields because the three canonical orientations initially shift in the same direction and remain comparatively close in frequency. At intermediate fields, increasing orientational dispersion produces a broad, low-intensity ensemble response. Above approximately 14 mT, the branches of the $|Q_{yz}\rangle \leftrightarrow |Q_{xz}\rangle$ transition begin to reconverge in frequency, leading to its re-emergence as a distinct feature in the powder average. Full orientational averaging of the calculated transitions produces a powder-averaged simulation (Fig. S3) that reproduces the overall experimental map, showing that the observed field-dependent structure arises from the orientational dispersion of the same quintet transitions identified at zero field.

To test whether additional quintet transitions become optically visible in this low-field regime, we recorded broad ODMR spectra up to 2.5 GHz at 10 and 18 mT (Fig. S4), which shows no additional resonances. Thus, throughout the investigated field range, the experimentally observed ODMR response remains dominated by the lower three-state quintet manifold. The 0-18 mT map therefore provides a direct experimental picture of how the optically detected quintet transitions evolve towards the ZFS-Zeeman crossover regime.

### Room-temperature coherent quintet control and high-field spin polarisation

To confirm the characteristic high-field spin polarisation of SF-derived $^5$TT and to probe coherent spin manipulation at room temperature conditions, we use transient and pulsed EPR at ~350 mT. Fig. 3a, b shows the room-temperature trEPR spectra of **Tc2** and **Tc6** in a polystyrene film (0.1wt%), averaged over the first 500 ns after photoexcitation. The trEPR spectra of the two oligomers are closely similar, consistent with the similarity observed by zero-field ODMR, and are reproduced by EasySpin simulations[29] of a strongly exchange-coupled triplet pair $^5$TT state with $S = 2$, using ZFS parameters $|D_Q| = 440$ MHz and $|E_Q| = 25$ MHz. These parameters closely match those obtained from zero-field ODMR, providing a consistent spin-Hamiltonian description of the quintet across the two field regimes. Measurements at 20 K and 80 K in frozen toluene give spectra of similar shape (Fig. S5), showing that the strongly coupled quintet state persists across the investigated temperature range. The $S = 2$ assignment is further supported by pulsed EPR nutation measurements (Fig. S6, Supplementary Note 4) and field-nutation sweep (Fig. S7) at 20 K. No independent ($S = 1$) triplet contribution is observed in these spectra, consistent with previous photophysical characterisation of **Tc2** and **Tc6**, which found no evidence for long-lived dissociated triplets.[27]

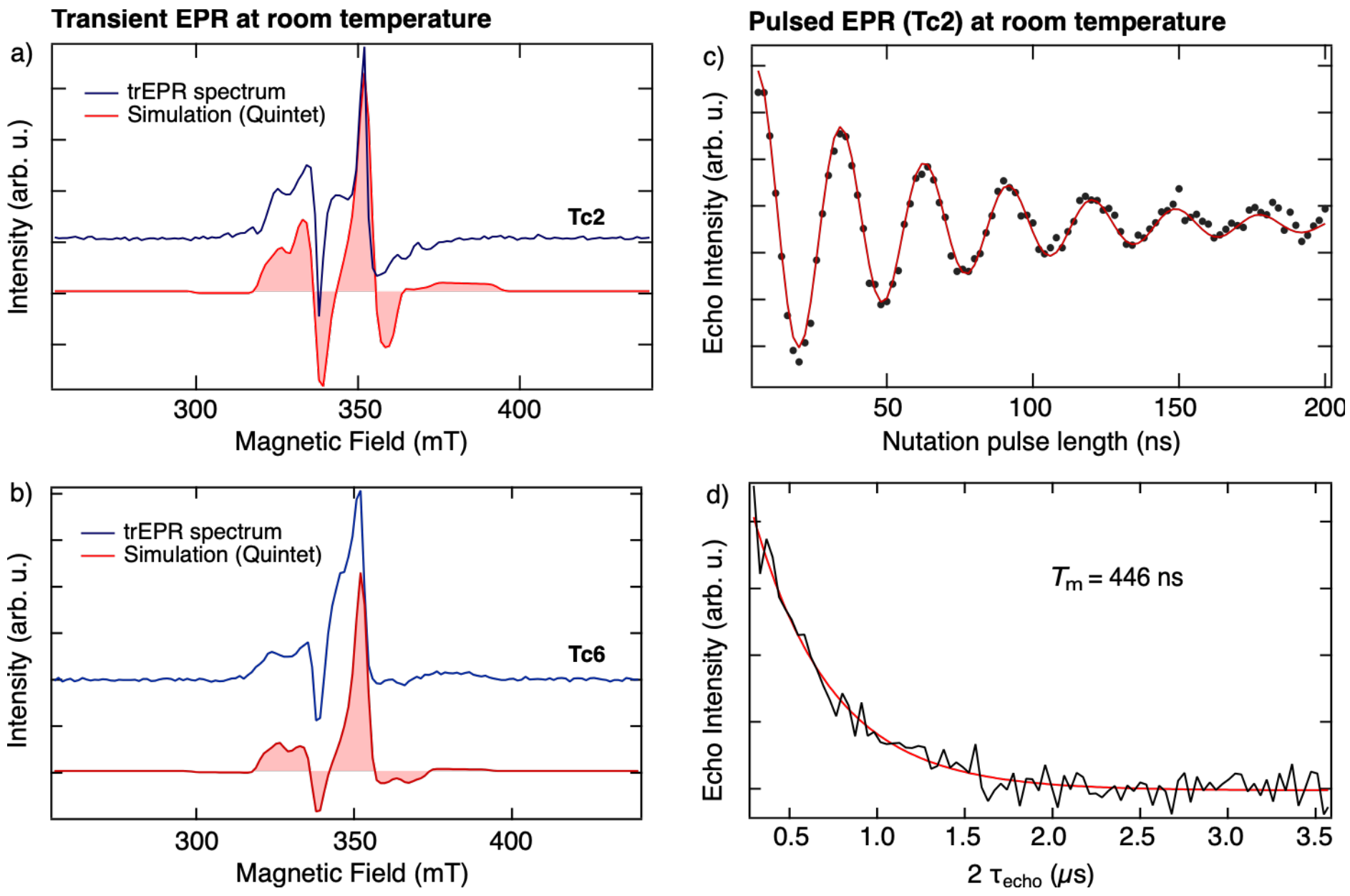

**Figure 3. Room-temperature quintet formation and coherent control.** a) Room-temperature trEPR spectrum of **Tc2**. b) Room-temperature trEPR of **Tc6**. Both spectra averaged over 0-500 ns after photoexcitation (blue), together with the corresponding EasySpin simulation for a strongly exchange-coupled quintet state ($S = 2$, red). c) Pulsed EPR on **Tc2** at room-temperature: Rabi oscillations at 352 mT, with damped-cosine fit (red), demonstrating coherent microwave control of the quintet. d) Room-temperature Hahn-echo decay of **Tc2** at 352 mT (black) with single-exponential fit (red), yielding a phase memory time $T_m = 446$ ns. All measurements were performed in polystyrene film containing 0.1 wt% of **Tc2**/**Tc6**.

Fig. 3c shows Rabi oscillations of **Tc2** recorded at 352 mT at $T = 293$ K, corresponding to coherently driving the $m_s = 0 \leftrightarrow \pm1$ quintet transition. The oscillation is well described by a damped cosine fit, yielding a Rabi frequency of 34.9 MHz for the applied microwave power. A Hahn-echo measurement at the same field gives a phase-memory time of $T_m = 446$ ns (Fig. 3d), while inversion-recovery measurements yield $T_1 = 2.7$ µs (Fig. S8). The room-temperature $T_m$ is comparable to those reported for structurally constrained quintet systems,[21,23,26] and demonstrates that substantial room-temperature coherence is retained despite the conformational degrees of freedom in these tetracene oligomers. Together, the Rabi oscillations and phase-memory measurement establish coherent microwave control of the SF-derived quintet at room temperature within the through-space coupled modular architecture.

The trEPR polarisation provides further insight into quintet formation at applied magnetic fields of ~350 mT. The population distribution obtained from the trEPR simulation is $p_{-2,-1,0,+1,+2} = [1, 0.30, 1, 0, 0.17]$ for **Tc2**, and similar for **Tc6** (Tab. S2), showing a strong population of $m_s = 0$ and $m_s = -2$ sublevels. Simultaneous population of $m_s = 0$ and $m_s = \pm2$ supports quintet formation through stochastic fluctuations of the interchromophore exchange interaction,[30,32] in contrast to the predominantly $m_s = 0$ polarisation expected for low-J singlet-quintet mixing[18,19]. The residual asymmetry between $m_s = +2$ and $m_s = -2$ is not captured by the ideal parallel-dimer model, indicating that additional effects associated with the non-parallel molecular geometry and spin dynamics contribute to the observed polarisation.[31] This interpretation is consistent with the non-parallel chromophore geometry indicated independently by the zero-field ODMR measurements. The alkyne-linked tetracene chromophores retain substantial torsional freedom within the HPB scaffold, providing a natural origin of fluctuations in the through-space exchange interaction.[33] Such torsional motion modulates the electronic overlap between neighbouring chromophores while retaining a finite through-space exchange interaction, consistent with the conditions required for stochastic quintet formation.

At both 80 K and room temperature, the trEPR spectrum undergoes further evolution on the microsecond timescale (Fig. S9). At room temperature, additional weak features develop on the microsecond timescale at the outer positions ~320 and ~370 mT and are more pronounced than at 80 K. Their enhanced late-time appearance is consistent with late-time loss of the strong spin correlation, similar to microsecond-timescale evolution reported for alkyne-linked non-conjugated pentacene dimers.[34]

## Discussion

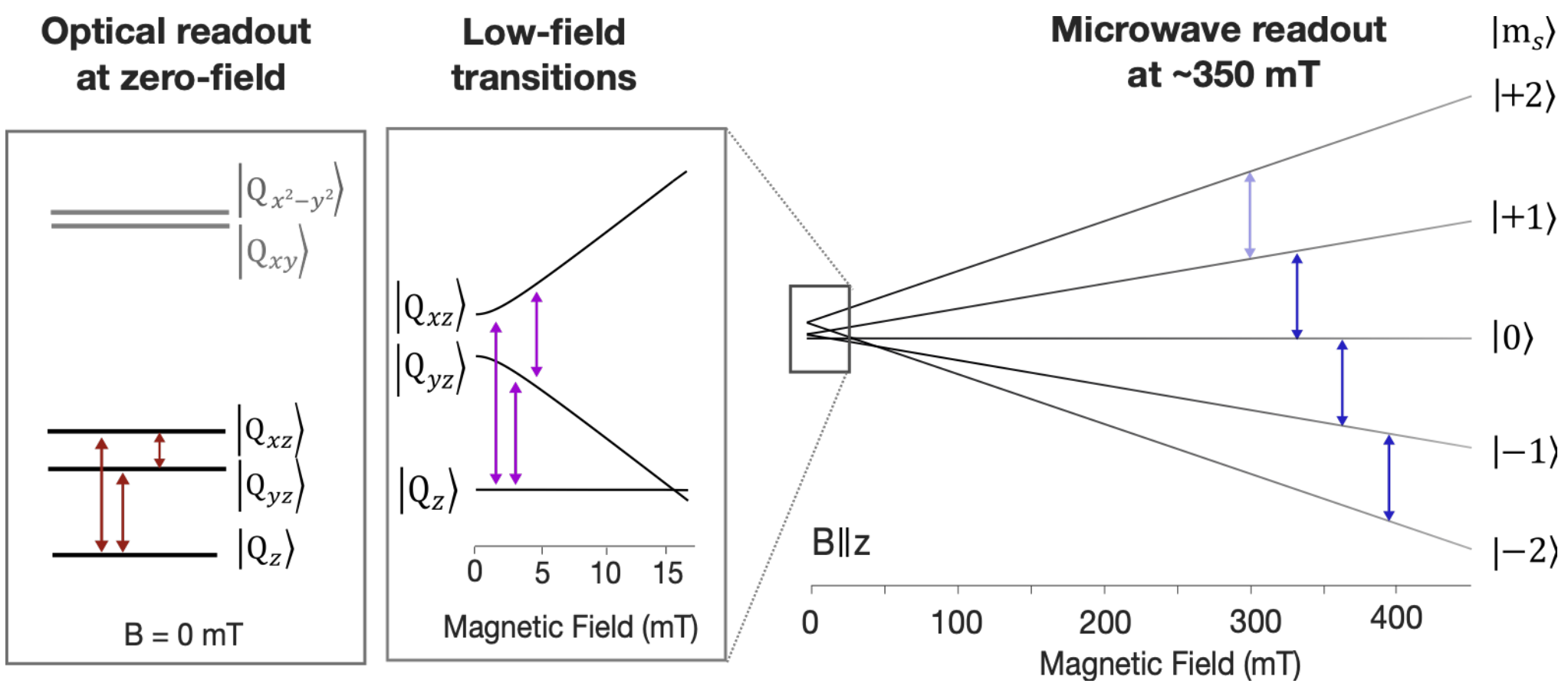

**Figure 4.** Field-dependent spin readout of SF-derived quintet states in **Tc2** and **Tc6**. Schematic overview of the complementary readout regimes explored in this work. Left: zero-field ODMR in the intrinsic quintet eigenbasis, with observed transitions in the lower three-state manifold. Centre: conceptual evolution of the optically addressable transitions in the low-field regime (0-18 mT). Right: Readout of microwave transitions in high-field (~350 mT) EPR in the Zeeman basis. Low-field and high-field levels plot by EasySpin with $S = 2$ and ZFS splitting parameters $|D_Q| = 440$ MHz, $|E_Q| = 30$ MHz, here shown for B||z.

These results establish that SF-derived quintets in through-space coupled tetracene oligomers can be optically addressed at room temperature without an applied magnetic field, respond strongly to weak magnetic fields, and retain coherent microwave control at room temperature. We also established that these properties can coexist in a chemically modular through-space coupled architecture under ambient conditions. Fig. 4 summarises the evolution of the quintet spin states and their experimental access from the intrinsic zero-field states, through the low-field regime, to the high-field Zeeman basis.

Zero-field ODMR (Fig. 1) directly resolves the spin transitions that are optically addressable in the absence of an applied magnetic field. The observed $D_Q \pm 3E_Q$ and $6E_Q$ resonances provide frequency-selective access to transitions within the zero-field states of the quintet without relying on Zeeman splitting. In particular, the $6E_Q$ transition demonstrates a non-equilibrium population difference between $|Q_{xz}\rangle$ and $|Q_{yz}\rangle$, enabled by the non-parallel chromophore geometry, providing an additional optically addressable transition beyond the ideal parallel formation pathways. This spectral separation allows individual two-level transitions within the five-level molecular qudit to be selectively addressed at different microwave frequencies, providing a basis for microwave pulse sequences that sequentially connect different pairs of quintet states. Importantly, the SF photophysics provide an intrinsic optical readout channel: the forward SF pathway generates the non-equilibrium $^5$TT population, while the reverse pathway reconnects the correlated triplet pair to the emissive singlet state. Microwave-driven redistribution of the quintet populations can therefore be detected directly through changes in photoluminescence, providing an optical interface to the intrinsic spin states. While we previously demonstrated this room temperature optical readout of quintets at X-band magnetic fields,[22] the present measurements show that the intrinsic Cartesian quintet states remain optically addressable at zero-field, connecting the field-free spin polarisation directly to the emissive SF pathway.

The low-field ODMR map (Fig. 2) directly follows this response as the Zeeman interaction is introduced and becomes comparable to the intrinsic quintet ZFS (Fig. 4). The resulting frequency-field behaviour is reproduced by orientationally averaged simulations using the same spin-Hamiltonian parameters, linking the zero-field and low-field responses within a common description. The different optically accessible transitions exhibit distinct frequency-field responses, establishing the low-field behaviour of the multilevel quintet and providing potential operating transitions for future molecular spin sensing. At higher magnetic fields, the same quintet evolves towards the Zeeman-defined basis probed by trEPR (Fig. 3). The $|Q_{xz}\rangle/|Q_{yz}\rangle$ states evolve towards the $m_s$ = ±1 manifold on introduction of a Zeeman interaction, whereas the high-field spectrum is dominated by $m_s$ = 0 and $m_s$ = -2 transitions (Fig. 4). The zero- and high-field measurements therefore resolve different spin polarisation: EPR reports the non-equilibrium polarisation in the Zeeman basis, whereas zero-field ODMR resolves population differences between the intrinsic Cartesian eigenstates. Furthermore, the optical detection of the microwave-induced changes in emission at zero-field provides a direct optical interface to the molecular spin state, which is advantageous for future molecular quantum communication.

The **Tc2**/**Tc6** hexaphenylbenzene-based architecture demonstrates that the molecular requirements for quintet formation and room-temperature coherence can be accommodated within a conformationally flexible through-space coupled framework. Previous photophysical studies established efficient intramolecular singlet fission in this architecture, with predominantly strongly coupled triplet pairs and no clear evidence for long-lived weakly coupled pairs or free triplets[27]. The present ODMR and EPR measurements show that these strongly bound pairs access a well-defined quintet state, while the closely matching ZFS parameters and spin polarisation of **Tc2** and **Tc6** indicate that the quintet remains local to a neighbouring tetracene pair as the oligomer is extended. Retaining this strongly correlated pair is advantageous for a molecular qudit because dissociation into independent triplets would remove the intrinsic multilevel $S$ = 2 spin manifold. The high-field spin polarisation further supports a stochastic exchange-fluctuation mechanism for quintet formation, consistent with the conformational degrees of freedom and through-space coupling of the non-parallel tetracene pair. The room-temperature phase-memory time of $T_m$ = 446 ns shows that substantial quintet coherence is nevertheless retained. This coexistence suggests that conformational dynamics can modulate the exchange interaction required for quintet formation without causing equally strong dephasing between the resulting quintet sublevels, expanding the molecular-design space beyond rigidly engineered chromophore arrangements.

## Conclusion

We demonstrate that SF-derived quintet states in through-space coupled tetracene oligomers can be optically initialised and read out at room temperature and without an applied magnetic field, providing direct optical access to spin transitions between the intrinsic zero-field quintet states. The distinct evolution of these optically detected transitions under weak magnetic fields further establishes field-responsive spin channels relevant to future molecular quantum sensing. Comparison with conventional EPR connects these field-free states to the characteristic spin polarisation of the SF-derived quintet in the high-field regime and supports quintet formation through stochastic exchange fluctuations. The quintet retains substantial room-temperature coherence, with a phase-memory time of 446 ns despite the conformational degrees of freedom of the molecular architecture, showing that the dynamics enabling high-spin formation can coexist with coherent spin manipulation. Through-space SF oligomers therefore provide a chemically modular molecular platform for strongly spin-polarized quintet formation, optical zero-field readout and room-temperature coherent control, offering a route towards field-free optical spin interfaces and molecular quantum sensing.

## Methods

### Sample preparation

**Tc2** and **Tc6** were synthesised as reported previously.[27] Toluene solutions (20-50 μM) were prepared under nitrogen atmosphere and subjected to three freeze-pump-thaw cycles before flame-sealing in EPR tubes for low-temperature measurements. For room-temperature ODMR and pulsed EPR, polystyrene films (0.1 wt%) were prepared by dispersing the compound in a polystyrene/toluene solution, drop-casting, and drying in inert atmosphere. EPR measurements in frozen solution and polystyrene films were compared at 80K to ensure same spin properties (Fig. S10). For ODMR, samples were encapsulated in inert atmosphere while for EPR thin-films, samples were flame-sealed under vacuum.

### Transient EPR

trEPR experiments were performed on a Bruker E680 EleXSys spectrometer using a Bruker ER4118-MD5-W1 dielectric resonator (9.72 GHz) in an Oxford Instruments CF935 cryostat (T = 20-80 K for frozen-solution measurements, room temperature for polystyrene thin-film measurements). Optical excitation was provided by an Ekspla NT230 laser (pulse duration 3 ns, repetition rate 50 Hz, pulse energy 1.0 mJ, $\lambda$ = 520 nm). Two-dimensional datasets were recorded in Bruker Xepr and spectral simulations were performed using EasySpin.[29] Comparison of thin-film samples and frozen solution samples at 80K confirmed the same configuration of the spin states in both sample architectures.

### Pulsed EPR

Rabi nutation experiments were performed on the same spectrometer as trEPR using rectangular microwave pulses (4-200 ns) at fixed magnetic fields of B = 326 mT and 355 mT (20 K) and B = 352 mT (room temperature, **Tc2** in polystyrene film). A delay after flash (DAF) of 200 ns was used. FFT analysis of the nutation signal yielded the nutation frequencies. Phase memory time at room temperature $T_m$ was extracted from Hahn-echo decays at 352 mT ($\pi$ = 32 ns, $\tau$ swept from 140 to 1928 ns in 12 ns steps), with the echo intensity fitted as a function of the total evolution time \(2\tau\) to a single-exponential decay. $T_1$ was extracted from echo-detected inversion-recovery measurements at 352 mT, and fitted to a single-exponential function.

**Zero-field and field-dependent ODMR**

ODMR was performed on a home-built zero-field molecular ensemble setup. Optical excitation was provided by a 532 nm CW laser (Laser Quantum Ventus 532). Photoluminescence was collected by a silicon photodiode (Hamamatsu) through a 550 nm long-pass filter, amplified by a transimpedance amplifier (Femto DHPCA-200), and detected with a lock-in amplifier (Stanford Research Systems SR830) referenced to on/off amplitude modulation of the microwave source (Stanford Research Systems SG384) at 413 Hz. The microwave frequency was swept from 0 to 2.5 GHz. For field-dependent measurements, a static magnetic field of 0-18 mT was applied via a small electromagnet and the microwave frequency was swept at each field step, generating the 2D map of Fig. 2a.

**Supporting Information**

Supporting Information contains additional ODMR measurements, transient and pulsed EPR measurements, EasySpin Simulations, temperature-dependent measurements, and supplementary discussion of zero-field transitions and Rabi nutation experiments.

## Acknowledgements

We acknowledge support from the UK Engineering and Physical Sciences Research Council (EPSRC), grant EP/Z534249/1. J. G. acknowledges support through a Research Fellowship from the Royal Commission for the Exhibition of 1851. P.G. and M. M. acknowledge financial support from the Polish National Science Centre (SONATA BIS: 2023/50/E/ST5/00263) and the Foundation for Polish Science (FirstTEAM: FENG.02.02-IP.05-0055/24). W.K.M. acknowledges EPSRC (grant to CAESR, EP/L011972/1) & John Fell Fund (0007019 & 0010710). H.L.S. acknowledges support from the Royal Society via Royal Society University Research Fellowship.

## References

1 Wasielewski, M. R.; Forbes, M. D. E.; Frank, N. L.; Kowalski, K.; Scholes, G. D.; Yuen-Zhou, J.; Baldo, M. A.; Freedman, D. E.; Goldsmith, R. H.; Goodson, T., III; et al. Exploiting chemistry and molecular systems for quantum information science. *Nat. Rev. Chem.* **2020**, 4, 490-504.

2 Tesiman, C. A.; Oxborrow, M.; Attwood, M. Surveying optically addressable spin qubits for quantum information and sensing technology. *npj Quantum Mater.* **2026**, *11*, 46.

3 Bonizzoni, C.; Ghirri, A.; Santanni, F.; Affronte, M. Quantum sensing of magnetic fields with molecular spins. *npj Quantum Inf.* **2024**, *10*, 41.

4 Weiss, L. R.; Smith, G. T.; Murphy, R. A.; Golesorkhi, B.; Méndez Méndez, J. A.; Patel, P.; Niklas, J.; Poluektov, O. G.; Long, J. R.; Awschalom, D. D. A high-resolution molecular spin-photon interface at telecommunication wavelengths. *Science* **2025**, 390, 76-81.

5 Gorgon, S.; Lv, K.; Grüne, J.; Drummond, B. H.; Myers, W. K.; Londi, G.; Ricci, G.; Valverde, D.; Tonnelé, C.; Murto, P.; *et al.* Reversible spin-optical interface in luminescent organic radicals. *Nature* **2023**, *620*, 538-544.

6 Smyser, K. E.; Eaves, J. D. Singlet fission for quantum information and quantum computing: the parallel JDE model. *Sci. Rep.* **2020**, *10*, 18480.

7 Lewis, S. G.; Smyser, K. E.; Eaves, J. D. Clock transitions guard against spin decoherence in singlet fission. *J. Chem. Phys.* **2021**, *155*, 194109.

8 Gruber, A.; Dräbenstedt, A.; Tietz, C.; Fleury, L.; Wrachtrup, J.; von Borczyskowski, C. Scanning confocal optical microscopy and magnetic resonance on single defect centers. *Science* **1997**, *276*, 2012-2014.

9 Doherty, M. W.; Manson, N. B.; Delaney, P.; Jelezko, F.; Wrachtrup, J.; Hollenberg, L. C. L. The nitrogen-vacancy colour centre in diamond. *Phys. Rep.* **2013**, *528*, 1-45.
10 Awschalom, D. D.; Hanson, R.; Wrachtrup, J.; Zhou, B. B. Quantum technologies with optically interfaced solid-state spins. *Nat. Photonics* **2018**, *12*, 516-527.
11 Balasubramanian, G.; Chan, I. Y.; Kolesov, R.; Al-Hmoud, M.; Tisler, J.; Shin, C.; Kim, C.; Wojcik, A.; Hemmer, P. R.; Krueger, A.; *et al.* Nanoscale imaging magnetometry with diamond spins under ambient conditions. *Nature* **2008**, *455*, 648-651.
12 Vetter, P. J.; Marshall, A.; Genov, G. T.; Weiss, T. F.; Striegler, N.; Großmann, E. F.; Oviedo-Casado, S.; Cerrillo, J.; Prior, J.; Neumann, P.; *et al.* Zero- and low-field sensing with nitrogen-vacancy centers. *Phys. Rev. Appl.* **2022**, *17*, 044028.
13 Bayliss, S. L.; Laorenza, D. W.; Mintun, P. J.; Kovos, B. D.; Freedman, D. E.; Awschalom, D. D. Optically addressable molecular spins for quantum information processing. *Science* **2020**, *370*, 1309-1312.
14 Mena, A.; Mann, S. K.; Cowley-Semple, A.; Bryan, E.; Heutz, S.; McCamey, D. R.; Attwood, M.; Bayliss, S. L. Room-temperature optically detected coherent control of molecular spins. *Phys. Rev. Lett.* **2024**, *133*, 120801.
15 Singh, H.; D'Souza, N.; Zhong, K.; Druga, E.; Oshiro, J.; Blankenship, B.; Montis, R.; Reimer, J. A.; Breeze, J. D.; Ajoy, A. Room-temperature quantum sensing with photoexcited triplet electrons in organic crystals. *Phys. Rev. Res.* **2025**, *7*, 013192.
16 Joshi, G.; Dill, R. D.; Thorley, K. J.; Anthony, J. E.; Reid, O. G.; Johnson, J. C. Optical readout of singlet fission biexcitons in a heteroacene with photoluminescence detected magnetic resonance. *J. Chem. Phys.* **2022**, *157*, 164702.
17 Smith, M. B.; Michl, J. Singlet fission. *Chem. Rev.* **2010**, *110*, 6891-6936.
18 Weiss, L. R.; Bayliss, S. L.; Kraffert, F.; Thorley, K. J.; Anthony, J. E.; Bittl, R.; Friend, R. H.; Rao, A.; Greenham, N. C.; Behrends, J. Strongly exchange-coupled triplet pairs in an organic semiconductor. *Nat. Phys.* **2017**, *13*, 176-181.
19 Tayebjee, M. J. Y.; Sanders, S. N.; Kumarasamy, E.; Campos, L. M.; Sfeir, M. Y.; McCamey, D. R. Quintet multiexciton dynamics in singlet fission. *Nat. Phys.* **2017**, *13*, 182-188.
20 Bayliss, S. L.; Weiss, L. R.; Kraffert, F.; Granger, D. B.; Anthony, J. E.; Behrends, J.; Bittl, R. Probing the wave function and dynamics of the quintet multiexciton state with coherent control in a singlet fission material. *Phys. Rev. X* **2020**, *10*, 021070.
21 Jacobberger, R. M.; Qiu, Y.; Williams, M. L.; Krzyaniak, M. D.; Wasielewski, M. R. Using molecular design to enhance the coherence time of quintet multiexcitons generated by singlet fission in single crystals. *J. Am. Chem. Soc.* **2022**, *144*, 2276-2283.
22 Grüne, J.; Montanaro, S.; Bradbury, T. W.; Sharma, A.; Dowland, S.; Gillett, A. J.; Gorgon, S.; Millington, O.; Myers, W. K.; Behrends, J.; *et al.* High-spin state dynamics and quintet-mediated emission in intramolecular singlet fission. *Nat. Commun.* **2026**, *17*, 777.
23 Ishii, W.; Psarra, M.; Fuki, M.; Smitran, M.; Ishibashi, K.; Parmar, B.; Inglis, A.; Pidgeon, J. P.; Sato, S.; Watanabe, G.; *et al.* Macrocyclic parallel rubrene dimers for room-temperature optically addressable quintet spin qudits. *ChemRxiv* **2026**.
24 MacDonald, T. S. C.; Tayebjee, M. J. Y.; Collins, M. I.; Kumarasamy, E.; Sanders, S. N.; Sfeir, M. Y.; Campos, L. M.; McCamey, D. R. Anisotropic multiexciton quintet and triplet dynamics in singlet fission via pulsed electron spin resonance. *J. Am. Chem. Soc.* **2023**, *145*, 15275-15283.
25 Paul, S.; Yampolskyy, O.; Wu, Z.; Müllen, K.; Basché, T. Singlet fission among two single molecules. *J. Am. Chem. Soc.* **2026**, *148*, 26343-26352.
26 Ishii, W.; Fuki, M.; Bu Ali, E. M.; Sato, S.; Parmar, B.; Yamauchi, A.; Mulyadi, C. H.; Uji, M.; Rivero, S. M.; Watanabe, G.; *et al.* Macrocyclic parallel dimer showing quantum coherence of quintet multiexcitons at room temperature. *J. Am. Chem. Soc.* **2024**, *146*, 25527-25535.
27 Majdecki, M.; Hsu, C.-H.; Wang, C.-H.; Shi, E. H.-C.; Zakrocka, M.; Wei, Y.-C.; Chen, B.-H.; Lu, C.-H.; Yang, S.-D.; Chou, P.-T.; *et al.* Singlet fission in a new series of systematically designed through-space coupled tetracene oligomers. *Angew. Chem. Int. Ed.* **2024**, *63*, e202401103.

28 Hasobe, T.; Nakamura, S.; Tkachenko, N. V.; Kobori, Y. Molecular design strategy for high-yield and long-lived individual doubled triplet excitons through intramolecular singlet fission. *ACS Energy Lett.* **2022**, *7*, 390-400.
29 Stoll, S.; Schweiger, A. EasySpin, a comprehensive software package for spectral simulation and analysis in EPR. *J. Magn. Reson.* **2006**, *178*, 42-55.
30 Collins, M. I.; McCamey, D. R.; Tayebjee, M. J. Y. Fluctuating exchange interactions enable quintet multiexciton formation in singlet fission. *J. Chem. Phys.* **2019**, *151*, 164104.
31 Dill, R. D.; Smyser, K. E.; Rugg, B. K.; Damrauer, N. H.; Eaves, J. D. Entangled spin-polarized excitons from singlet fission in a rigid dimer. *Nat. Commun.* **2023**, *14*, 1180.
32 Collins, M. I.; Campaioli, F.; Tayebjee, M. J. Y.; Cole, J. H.; McCamey, D. R. Quintet formation, exchange fluctuations, and the role of stochastic resonance in singlet fission. *Commun. Phys.* **2023**, *6*, 64.
33 Kobori, Y.; Fuki, M.; Nakamura, S.; Hasobe, T. Geometries and terahertz motions driving quintet multiexcitons and ultimate triplet–triplet dissociations via the intramolecular singlet fissions. *J. Phys. Chem. B* **2020**, *124*, 9411-9419.
34 Basel, B. S.; Zirzlmeier, J.; Hetzer, C.; Phelan, B. T.; Krzyaniak, M. D.; Reddy, S. R.; Coto, P. B.; Horwitz, N. E.; Young, R. M.; White, F. J.; *et al.* Unified model for singlet fission within a non-conjugated covalent pentacene dimer. *Nat. Commun.* **2017**, *8*, 15171.